\documentclass[11pt]{article}
\usepackage{amsmath,amsthm,bm, amscd}
\usepackage{fullpage}
\usepackage[nofullpage,full,titlepage,nousetoc,nouselot,nouselof,hylinks,final]{boaz}

\usepackage{color}

\usepackage{graphicx}

\newcommand{\supp}{\mathsf{supp}}

\newcommand{\twext}{\mathsf{2Ext}}
\newcommand{\bitext}{\mathsf{BitExt}}
\newcommand{\bfext}{\mathsf{BFExt}}

\newcommand{\zo}{\bits}

\newcommand{\bq}{\overline{Q}}
\newcommand{\bi}{\mathbf{I}}

\theoremstyle{definition}

\newcommand{\eps}{\epsilon}

\newcommand{\ac}{\mathsf{AC^0}}

\newcommand{\Ext}{\mathsf{Ext}}

\newcommand{\BI}{\begin{itemize}}
\newcommand{\EI}{\end{itemize}}
\newcommand{\BE}{\begin{enumerate}}
\newcommand{\EE}{\end{enumerate}}

\newtheorem{thm}{Theorem}      
\newcommand{\BT}{\begin{theorem}}   \newcommand{\ET}{\end{theorem}}
\newcommand{\BD}{\begin{definition}}   \newcommand{\ED}{\end{definition}}
\newcommand{\BCR}{\begin{corollary}} \newcommand{\ECR}{\end{corollary}}
\newtheorem{constr}[thm]{Construction}
\newcommand{\BCT}{\begin{constr}} \newcommand{\ECT}{\end{constr}}
\newcommand{\BL}{\begin{lemma}}   \newcommand{\EL}{\end{lemma}}

\newcommand{\BP}{\begin{proposition}}   \newcommand{\EP}{\end{proposition}}
\newcommand{\BCM}{\begin{claim}}   \newcommand{\ECM}{\end{claim}}
\newcommand{\BF}{\begin{fact}}   \newcommand{\EF}{\end{fact}}
\newcommand{\BA}{\begin{assumption}}   \newcommand{\EA}{\end{assumption}}
\newcommand{\tabincell}[2]{\begin{tabular}{@{}#1@{}}#2\end{tabular}}

\def\eps{\varepsilon}

\makeatletter
\def\ExtendSymbol#1#2#3#4#5{\ext@arrow 0099{\arrowfill@#1#2#3}{#4}{#5}}
\def\RightExtendSymbol#1#2#3#4#5{\ext@arrow 0359{\arrowfill@#1#2#3}{#4}{#5}}
\def\LeftExtendSymbol#1#2#3#4#5{\ext@arrow 6095{\arrowfill@#1#2#3}{#4}{#5}}
\makeatother

\begin{document}

\begin{titlepage}
\def\thepage{}

\date{}
\title{
Improved Constructions of Two-Source Extractors
}

\author{
Xin Li\\
Department of Computer Science\\
Johns Hopkins University\\
Baltimore, MD 21218, U.S.A.\\
lixints@cs.jhu.edu
}

\maketitle \thispagestyle{empty}

\begin{abstract}
In a recent breakthrough \cite{CZ15}, Chattopadhyay and Zuckerman gave an explicit two-source extractor for min-entropy $k \geq \log^C n$ for some large enough constant $C$. However, their extractor only outputs one bit. In this paper, we improve the output of the two-source extractor to $k^{\Omega(1)}$, while the error remains $n^{-\Omega(1)}$.

Our improvement is obtained by giving a better extractor for $(q, t, \gamma)$ non-oblivious bit-fixing sources, which can output $t^{\Omega(1)}$ bits instead of one bit as in \cite{CZ15}.
\end{abstract}
\end{titlepage}

\section{Introduction}
In theoretical computer science and in particular the area of pseudorandomness, one deals with the problem of either reducing the number of random bits used in applications (ideally, completely removing the use of random bits) or replacing the uniform random bits needed by random sources with very weak quality. The former is done by constructing \emph{pseudorandom generators} and the latter is done by constructing \emph{randomness extractors}. Both objects have been the focus of extensive study over the past several decades. 

In this paper, we focus on randomness extractors and in particular we study the problem of constructing randomness extractors for independent general weak random sources, where a weak random source is modeled by a distribution with a certain amount of entropy. 

\begin{definition}
The \emph{min-entropy} of a random variable~$X$ is
\[ H_\infty(X)=\min_{x \in \supp(X)}\log_2(1/\Pr[X=x]).\]
For $X \in \zo^n$, we call $X$ an $(n,H_\infty(X))$-source, and we say $X$ has
\emph{entropy rate} $H_\infty(X)/n$.
\end{definition}

A well known result is that with just one weak source as input, no deterministic extractor can work for all $(n, k)$ sources even when $k=n-1$. Due to this, there have been two different directions and relaxations for randomness extractors. The first one is to give the extractor an additional independent uniform random seed, which is generally much shorter than the source (e.g., $O(\log n))$. These extractors are called \emph{seeded extractors} and were introduced by Nisan and Zuckerman \cite{NisanZ96}. 

\begin{definition}(Seeded Extractor)\label{def:strongext}
A function $\Ext : \bits^n \times \bits^d \rightarrow \bits^m$ is  a \emph{$(k,\eps)$-extractor} if for every source $X$ with min-entropy $k$
and independent $Y$ which is uniform on $\zo^d$,
\[|\Ext(X, Y)-U_m | \leq \e.\]
If in addition we have $|(\Ext(X, Y), Y) - (U_m, Y)| \leq \e$ then we say it is a \emph{strong $(k,\eps)$-extractor}.
\end{definition}

With the help of the extra seed, it is now indeed possible to construct extractors for all weak random sources. Seeded extractors have found numerous applications in theoretical computer science, and today we have nearly optimal constructions of such extractors (e.g., \cite{LuRVW03, GuruswamiUV09, DvirW08, DvirKSS09}).

In the other direction, one considers building deterministic extractors for sources which have special structures. In this case, one important class of sources is the so called \emph{independent sources}. Here, the extractor is given as input more than one general weak random sources, and the sources are independent each other. Using the probabilistic method, one can show that there exists a deterministic extractor for just two independent sources with logarithmic min-entropy, which is optimal since extractors for one weak source do not exist. In fact, the probabilistic method shows that with high probability a random function is such a two-source extractor. However, the most interesting and important part is to give explicit constructions of such functions, which turns out to be highly challenging. 

The first explicit construction of a two-source extractor appeared in \cite{ChorG88}, where Chor and Goldreich showed that the well known Lindsey's lemma gives an extractor for two independent $(n, k)$ sources with $k > n/2$. Since then there has been essentially no progress on two-source extractors until in 2005 Bourgain \cite{Bourgain05} gave a construction that breaks the entropy rate $1/2$ barrier, and works for two independent $(n, 0.49n)$ sources. In a different work, Raz \cite{Raz05} gave an incomparable result of two source extractors which requires one source to have min-entropy larger than $n/2$, while the other source can have min-entropy $O(\log n)$.

Given the difficulty of constructing explicit two-source extractors, much research has been focusing on a slightly more general model, where the extractor is allowed to have more then two independent sources as the input. Starting from \cite{BarakIW04}, there has been a long line of fruitful results \cite{BarakIW04, Raz05, Bourgain05, Rao06, BarakRSW06, Li11b, Li13a, Li13b, Li15, Cohen15}, which introduced many new techniques and culminated in the three source extractor of exponentially small error by the author \cite{Li15}. However, in the two source case the situation has not been improved.

Recently, Chattopadhyay and Zuckerman \cite{CZ15} made an exciting breakthrough to the problem of constructing explicit two-source extractors. They gave an explicit two-source extractors for $(n, k)$ sources with $k \geq \log^C n$ for some large enough constant $C$. This dramatically improves the situation of two-source extractors and is actually near optimal. However, their construction only outputs one bit.

In this paper, we improve the output length of the two-source extractor in \cite{CZ15} to $k^{\Omega(1)}$. Specifically, we have the following theorem.

\BT \emph{(Main theorem)}. 
There exists a constant $C>0$ such that for all $n \in \N$, there exists a polynomial time computable function $\twext: \bits^n \times \bits^n \to \bits^m$ satisfying the following: if $X, Y$ are two independent $(n, k)$ sources with $k \geq \log^C n$, then 

\[|(\twext(X, Y), Y)-(U_m, Y)| \leq \e,\]
where $m=k^{\Omega(1)}$ and $\e=n^{-\Omega(1)}$.
\ET

Since the extractor is strong in $Y$, if we don't need a strong extractor, then we can use the output of $\twext$ to extract from $Y$ and output almost all the min-entropy. For example, we have the following theorem.

\BT 
There exists a constant $C>0$ such that for all $n \in \N$, there exists a polynomial time computable function $\twext: \bits^n \times \bits^n \to \bits^m$ satisfying the following: if $X, Y$ are two independent $(n, k)$ sources with $k \geq \log^C n$, then 

\[|\twext(X, Y)-U_m| \leq \e,\]
where $m=0.9k$ and $\e=n^{-\Omega(1)}$.
\ET

\subsection{Non-Oblivious Bit-Fixing Sources}
As in \cite{CZ15}, our construction is also based on reducing two independent sources to a special kind of non-oblivious bit-fixing source. Following \cite{CZ15}, we formally define such sources.

\BD
A distribution $\cal D$ on n bits is t-wise independent if the restriction of $\cal D$ to any $t$ bits is uniform. Further $\cal D$ is a $(t, \e)$-wise independent distribution if the distribution obtained by restricting $\cal D$ to any $t$ coordinates is $\e$-close to uniform.
\ED

\BD
A source $X$ on $\bits^n$ is called a $(q,t)$-non-oblivious bit-fixing source if there exists a subset of coordinates $Q \subseteq [n]$ of size at most $q$ such that the joint distribution of the bits indexed by $\overline{Q} = [n] \setminus Q$ is $t$-wise independent. The bits in the coordinates indexed by $Q$ are allowed to arbitrarily depend on the bits in the coordinates indexed by $\overline{Q}$.

If the joint distribution of the bits indexed by $\overline{Q}$ is $(t,\gamma)$-wise independent then $X$ is said to be a $(q, t, \gamma)$-non-oblivious bit-fixing source.
\ED

Our main theorem of the improved two-source extractor actually follows directly from our improvement of the extractor for a $(q, t, \gamma)$-non-oblivious bit-fixing source in \cite{CZ15}. Specifically, we have the following theorem.

\BT 
There exists a constant $c$ such that for any constant $\delta>0$ and all $n \in \N$, there exists an explicit extractor $\bfext: \bits^n \to \bits^m$ such that for any $(q, t, \gamma)$ non-oblivious bit-fixing source $X$ on $n$ bits with $q \leq n^{1-\delta}$, $t \geq c \log^{21} n$ and $\gamma \leq 1/n^{t+1}$, we have that
\[|\bfext(X)-U_m| \leq \e,\]

where $m=t^{\Omega(1)}$ and $\e=n^{-\Omega(1)}$.
\ET

\tableref{table:result} summarizes our results compared to previous constructions of independent source extractors. Note that in the table the output length are all under the conditions of being strong extractors. 

\begin{table}[ht] 
\centering 
\begin{tabular}{|c|c|c|c|c|} 
\hline Construction & Number of Sources &  Min-Entropy &  Output & Error\\ 
\hline \cite{ChorG88} & 2 & $k \geq (1/2+\delta) n$, any constant $\delta$ & $\Theta(n)$ & $2^{-\Omega(n)}$ \\
\hline \cite{BarakIW04} & $\poly(1/\delta)$ & $\delta n$, any constant $\delta$ & $\Theta(n)$ & $2^{-\Omega(n)}$ \\
\hline \cite{BarakKSSW05} & 3 & $\delta n$, any constant $\delta$ & $\Theta(1)$& $O(1)$ \\
\hline  \cite{Raz05} & 3 & \tabincell{l}{One source: $\delta n$, any constant $\delta$. Other\\ sources may have $k \geq \polylog(n)$.} & $\Theta(1)$& $O(1)$ \\
\hline  \cite{Raz05}& 2 & \tabincell{l}{One source: $(1/2+\delta) n$, any constant $\delta$. \\Other source may have $k \geq \polylog(n)$} & $\Theta(k)$& $2^{-\Omega(k)}$ \\
\hline \cite{Bourgain05} & 2 &  \tabincell{l}{$(1/2-\alpha_0) n$ for some small universal \\constant $\alpha_0>0$} & $\Theta(n)$& $2^{-\Omega(n)}$ \\
\hline \cite{Rao06} & 3 & \tabincell{l}{One source: $\delta n$, any constant $\delta$. Other\\ sources may have $k \geq \polylog(n)$.} & $\Theta(k)$ & $2^{-k^{\Omega(1)}}$ \\
\hline \cite{Rao06} & $O(\log n /\log k)$ & $k \geq \polylog(n)$ & $\Theta(k)$ & $k^{-\Omega(1)}$ \\
\hline \cite{BarakRSW06} & $O(\log n /\log k)$ & $k \geq \polylog(n)$ & $\Theta(k)$ & $2^{-k^{\Omega(1)}}$ \\
\hline \cite{Li11b} & 3 & \tabincell{l}{$k=n^{1/2+\delta}$, any constant $\delta$}  & $\Theta(k)$& $k^{-\Omega(1)}$ \\
\hline \cite{Li13a} & $O(\log (\frac{\log n}{\log k}))+O(1)$ & $k \geq \polylog(n)$ & $\Theta(k)$ & $k^{-\Omega(1)}$ \\
\hline \cite{Li13b} & \tabincell{l}{$O(\frac{1}{\eta})+O(1),$ \\ $O(1)$ can be large} & $k \geq \log^{2+\eta} n$ & $\Theta(k)$ & \tabincell{l}{$n^{-\Omega(1)}+$ \\$2^{-k^{\Omega(1)}}$}  \\
\hline \cite{Li15} & 3 & $k \geq \log^{12} n$ & $\Theta(k)$ &  $2^{-k^{\Omega(1)}}$  \\
\hline \cite{Li15} & $\lceil \frac{14}{\eta} \rceil+2$ & $k \geq \log^{2+\eta} n$ & $\Theta(k)$ &  $2^{-k^{\Omega(1)}}$  \\
\hline \cite{Cohen15} & $3$ & $ \delta n, O(\log n), O(\log \log n)$ & $\Theta(\log n)$ &  $(\log n)^{-\Omega(1)}$  \\
\hline \cite{CZ15} & $2$ & $k \geq \log^C n$ for some large constant $C$. & $1$ &  $n^{-\Omega(1)}$  \\
\hline This work & $2$ & $k \geq \log^C n$ for some large constant $C$. & $k^{\Omega(1)}$ &  $n^{-\Omega(1)}$  \\
 \hline
\end{tabular}
\caption{\textbf{Summary of Results on Extractors for Independent Sources.}} 
\label{table:result}
\end{table}



\subsection{Overview of The Constructions and Techniques}
Here we give a brief overview of our constructions and the techniques. We first describe our new extractor for the $(q, t, \gamma)$ non-oblivious bit-fixing source on $n$ bits with $q \leq n^{1-\delta}$ for some constant $\delta>0$, and $\gamma \leq 1/n^{t+1}$. Our starting point is the one-bit deterministic extractor for such sources in \cite{CZ15}, which we'll call $\bitext$. We note that from the construction of \cite{CZ15}, (by setting the parameters appropriately) this function has the following properties. First, it is a depth-4 $\mathsf{AC^0}$ circuit with size $n^{O(1)}$. Second, since it's an extractor, for any $(q, t, \gamma)$ non-oblivious bit-fixing source $X$, we have $\bitext(X)$ is $n^{-\Omega(1)}$-close to uniform. Third, it's a resilient function, in the sense that any coalition of any $q$ bits has influence at most $q/n^{1-\frac{\delta}{2}}$.

We now describe how to extract more than one bit. One natural idea is to divide the source $X$ into many blocks and then apply $\bitext$ to each block. Indeed this is our first step. In the source $X$, we denote the ``bad bits" by $Q$, and the ``good bits" by $\bq$. To ensure that no block consists of only bad bits, we will divide $X$ into $n^{\alpha}$ blocks for some constant $\alpha  < \delta$ (it suffices to take $\alpha=\delta/4$). Thus we get $\ell=n^{\alpha}$ blocks $\{X_i, i \in [\ell]\}$ with each block containing $n'=n^{1-\alpha}$ bits. We now apply $\bitext$ to each block to obtain a bit $Y_i$. Of course, we will set up the parameters such that $\bitext$ is an extractor for $(q, t, \gamma)$ non-oblivious bit-fixing source on $n^{1-\alpha}$ bits.  

Now consider any block. Our observation is that since each block can contain at most $q \leq n^{1-\delta}$ bits from $Q$, the coalition of the bad bits in this block still has small influence. In particular, by a simple calculation shows that $q < n'^{1-\frac{3\delta}{4}}$ and thus for each block the influence of the bad bits is bounded by $q/n'^{1-\frac{3\delta}{8}}< n^{-\frac{3\delta}{8}}$. This means that with probability at least $1-n^{-\frac{3\delta}{8}}$ over the fixing of $X_i \cap \overline{Q}$, we have that $Y_i$ is fixed. Thus, by a simple union bound, with probability at least $1-n^{\alpha}n^{-\frac{3\delta}{8}}=1-n^{-\frac{\delta}{8}}$ over the fixing of $\bq$, we have that all $\{Y_i, i \in [\ell]\}$ are fixed. 

Now consider another distribution $X'$, which has the same distribution as $X$ for the bits in $\bq$, while the bits in $Q$ are fixed to $0$ independent of the bits in $\bq$. We let $Y'_i, i \in [\ell]$ be the corresponding $Y_i$'s obtained from $X'$ instead of $X$. By the above argument, with probability at least $1-n^{-\frac{\delta}{8}}$ over the fixing of $\bq$, $\{Y_i\}$ and $\{Y'_i\}$ are the same. Thus the joint distribution of $\{Y_i\}$ and $\{Y'_i\}$ are within statistical distance $n^{-\frac{\delta}{8}}$. Moreover, the bits in $\bq$ are $(t, \gamma)$-wise independent and thus they are $n^t \gamma \leq 1/n$-close to a truly $t$-wise independent distribution. From now on we will treat $\bq$ as being truly $t$-wise independent, since this only adds $1/n$ to the final error.

We will now choose a parameter $m=t^{\Omega(1)}$ for the output length. In addition, we take the generating matrix $G$ of an asymptotically good linear binary code with message length $m$, codeword length $r=O(m)$ and distance $d=\Omega(m)$. It is well known how to construct such codes (and thus the generating matrix) explicitly. Note that $G$ is an $m \times r$ matrix and any codeword can be generated by $w=vG$ for some vector $v \in \bits^m$, where all operations are in $\F_2$. We choose $m$ so that $r=O(m) \leq \ell$ and now we let $Y=(Y_1, \cdots, Y_r)$ be the random vector in $\F^r_2$. Similarly, we have $Y'=(Y'_1, \cdots, Y'_r)$. The output of our extractor will now be $Z=(Z_1, \cdots, Z_m)=GY$, where all operations are in $\F_2$.  

For the analysis let us consider $Z'=(Z'_1, \cdots, Z'_m)=GY'$. We will show that $Z'$ is close to uniform and then it follows that $Z$ is also close to uniform since they are within statistical distance $n^{-\Omega(1)}$ (since they are deterministic functions of $Y$ and $Y'$ respectively). To show this, we will use the XOR lemma. Consider any non-empty subset $S \subseteq [m]$ and $V'_S=\bigoplus_{i \in S} Z'_i$. Note that this is just $(\sum_{i \in S} G_i) Y'$ where $G_i$ stands for the $i$'th row of $G$. Note that $\sum_{i \in S} G_i$ is a codeword and thus has at least $d=\Omega(m)$ $1$'s. On the other hand, it can have at most $r=O(m)$ $1$'s. 

Note that the parity of up to $r$ bits can be computer by a depth-2 $\ac$ circuit of size $2^{O(r)}=2^{O(m)}$. Recall that each input bit $Y_i$ can be computed by a depth-4 $\ac$ circuit of size $n^{O(1)}$. Thus we see that each $V'_S$ can be computed by a depth-6 $\ac$ circuit of size at most $2^{O(m)}n^{O(1)}=2^{O(m)}$ if we choose $m > \log n$. \footnote{We can get rid of the intermediate negation gates with only a constant factor of blow-up in the circuit size, by standard tricks.} Note that all bits in $Q$ are fixed to $0$. Thus the inputs of the circuits are only from $\bq$. 

Now our goal is to ensure that $V'_S$ can be fooled by $t$-wise independent distributions with error $\e=2^{-m}$. By the theorem of Braverman \cite{Braverman10} and Tal \cite{Tal14}, it suffices to take $t=O(\log (2^{O(m)}/\e)^{21})=O(m^{21})$. Thus we can take $m=\Omega(t^{\frac{1}{21}})$. On the other hand, if $\bq$ are the uniform distribution, then $V'_S$ is the XOR of at least $d=\Omega(m)$ independent random variables, with each being $n^{-\Omega(1)}$-close to uniform. Thus in this case $V'_S$ is $(n^{-\Omega(1)})^d=2^{-\Omega(m\log n)}$-close to uniform. Together this means that $V'_S$ is $2^{-\Omega(m\log n)}+2^{-m}< 2^{1-m}$ close to uniform. Since this is true for any non-empty subset $S$, by a standard XOR lemma it now follows that $Z'$ is $2^{-\Omega(m)}$-close to uniform. Adding back the errors we see that $Z$ is $n^{-\Omega(1)}$-close to uniform.

Applying the reduction from two independent sources to a non-oblivious bit-fixing source, we immediately obtain our improved two-source extractor.\\

\noindent{\bf Organization.}
The rest of the paper is organized as follows.\ We give some preliminaries in \sectionref{sec:prelim}. We present our main construction of extractors in \sectionref{sec:ext}, and we conclude with some discussions and open problems in \sectionref{sec:conc}.

\section{Preliminaries} \label{sec:prelim}
We often use capital letters for random variables and corresponding small letters for their instantiations. Let $|S|$ denote the cardinality of the set~$S$. For $\ell$ a positive integer,
$U_\ell$ denotes the uniform distribution on $\zo^\ell$. When used as a component in a vector, each $U_\ell$ is assumed independent of the other components.
All logarithms are to the base 2.

\subsection{Probability distributions}
\begin{definition} [statistical distance]Let $W$ and $Z$ be two distributions on
a set $S$. Their \emph{statistical distance} (variation distance) is
\begin{align*}
\Delta(W,Z) \eqdef \max_{T \subseteq S}(|W(T) - Z(T)|) = \frac{1}{2}
\sum_{s \in S}|W(s)-Z(s)|.
\end{align*}
\end{definition}

We say $W$ is $\eps$-close to $Z$, denoted $W \approx_\eps Z$, if $\Delta(W,Z) \leq \eps$.
For a distribution $D$ on a set $S$ and a function $h:S \to T$, let $h(D)$ denote the distribution on $T$ induced by choosing $x$ according to $D$ and outputting $h(x)$.

\subsection{Influence of variables}
Following \cite{CZ15}, we define the influence of variables.

\BD
Let $f: \bits^n \to \bits$ be any boolean function on variables $x_1, \cdots, x_n$. The influence of a set $Q \subseteq \{x_1, \cdots, x_n\}$ on $f$, denoted by $\bi_Q  (f)$, is defined to be the probability that $f$ is undetermined after fixing the variables outside $Q$ uniformly at random. Further, for any integer $q$ define $\bi_q(f)=max_{Q \subseteq \{x_1, \cdots, x_n\}, |Q|=q} \bi_Q(f)$.

More generally, let $\bi_{Q, D}(f)$ denote the probability that $f$ is undermined when the variables outside $Q$ are fixed by sampling from the distribution $D$. We define $\bi_{Q, t} (f)=max_{D \in D_k} \bi_{Q, D} (f)$, where $D_t$ is the rest of all $t$-wise independent distributions. Similarly, $\bi_{Q, t, \gamma} (f)=max_{D \in D_{t, \gamma}} \bi_{Q, D} (f)$ where $D_{t, \gamma}$ is the set of all $(t, \gamma)$-wise independent distributions. Finally, for any integer $q$ define $\bi_{q, t}(f)=max_{Q \subseteq \{x_1, \cdots, x_n\}, |Q|=q} \bi_{Q, t}(f)$ and $\bi_{q, t, \gamma}(f)=max_{Q \subseteq \{x_1, \cdots, x_n\}, |Q|=q} \bi_{Q, t, \gamma}(f)$
\ED

\subsection{Prerequisites from previous work}

\BL [\cite{BarakIW04}] \label{lem:derror} 
Assume that $Y_1, Y_2, \cdots, Y_t$ are independent random variables over $\bits^n$ such that for any $i, 1 \leq i \leq t$, we have $|Y_i-U_n| \leq \e$. Let $Z=\oplus_{i=1}^{t} Y_i$. Then $|Z-U_n| \leq \e^t$.
\EL

To prove our construction is an extractor, we need the following definition and lemma.

\BD($\e$-biased space)
A random variable $Z$ over $\bits$ is $\e$-biased if $|\Pr[Z=0]-\Pr[Z=1]| \leq \e$. A sequence of 0-1 random variables $Z_1, \cdots, Z_m$ is \emph{$\e$-biased for linear tests} if for any nonempty set $S \subset \{1, \cdots, m\}$, the random variable $Z_S = \bigoplus_{i \in S} Z_i$ is $\e$-biased. 

\ED

The following lemma is due to Vazirani. For a proof see for example \cite{Oded95}

\BL \label{lem:vxor}
Let $Z_1, \cdots, Z_m$ be 0-1 random variables that are $\e$-biased for linear tests. Then, the distribution of $(Z_1, \cdots, Z_m)$ is $\e \cdot 2^{m/2}$-close to uniform.
\EL

\section{The Constructions of Extractors}\label{sec:ext}
In this section we describe our deterministic extractor for an $(q, t, \gamma)$-non-oblivious bit-fixing source on $n$ bits. We rely on the following result from \cite{CZ15}.

\BT [\cite{CZ15}] \label{thm:bitext}
There exists a constant $c>0$ such that for any $\delta>0$ and every large enough $n \in \N$ the following is true. Let $X$ be a $(q, t, \gamma)$ non-oblivious bit-fixing source on $n$ bits with $q \leq n^{1-\delta}$, $t \geq c \log^{18} n$ and $\gamma \leq 1/n^{t+1}$.  There exists a polynomial time computable monotone boolean function $\bitext: \bits^n \to \bits$ satisfying: 

\BI
\item $\bitext$ is a depth $4$ circuit in $\mathsf{AC}^0$ of size $n^{O(1)}$.

\item $|\mathsf{E}_{x \leftarrow X}[\bitext(x)]-\frac{1}{2}| \leq \frac{1}{n^{\Omega(1)}}.$

\item For any $q>0$, $\mathbf{I}_{q, t, \gamma}(\bitext) \leq q/n^{1-\frac{\delta}{2}}.$
\EI
\ET

We need the following result by Braverman \cite{Braverman10} and Tal \cite{Tal14} about fooling $\mathsf{AC}^0$ circuits with $t$-wise independent distributions.

\BT [\cite{Braverman10, Tal14}] \label{thm:ac0} 
Let $\cal D$ be any $t=t(m,d,\e)$-wise independent distribution on $\bits^n$. Then for any circuit $\cal C \in \mathsf{AC^0}$ of depth $d$ and size $m$,

\[|E_{x \sim U_n}[{\cal C} (x)]-E_{x \sim \cal D}[{\cal C} (x)]| \leq \e,\]
where $t(m,d,\e)=O(\log(m/\e))^{3d+3}$.
\ET

\BT [\cite{AGM03}] \label{thm:twise}
Let $\cal D$ be a $(t, \gamma)$-wise independent distribution on $\bits^n$. Then there exists a $t$-wise independent distribution on $\bits^n$ that is $n^t \gamma$-close to $\cal D$.
\ET

We also need an explicit asymptotically good binary linear codes:

\BD
A linear binary code of length $n$ and rank $k$ is a linear subspace $C$ with dimension $k$ of the vector space $\F^n_2$. If the distance of the code $C$ is $d$ we say that $C$ is an $[n,k,d]_2$ code. $C$ is asymptotically good if there exist constants $0< \delta_1, \delta_2<1$ s.t. $k \geq \delta_1 n$ and $d \geq \delta_2 n$.
\ED

Note that every linear binary code has an associated generating matrix $G \in \F^{k \times n}_2$, and every codeword can be expressed as $vG$, for some vector $v \in \F^k_2$.

It is well known that we have explicit constructions of asymptotically good binary linear code. For example, the Justensen codes constructed in \cite{Justensen72}.

Now we have the following construction and theorem:

We now present our extractor for a $(q, t, \gamma)$ non-oblivious bit-fixing source.

\begin{algorithmsub} 
{$\bfext(X)$ }
{$X$--- a $(q, t, \gamma)$ non-oblivious bit-fixing source on $n$ bits with $q \leq n^{1-\delta}$, $t \geq c \log^{21} n$ and $\gamma \leq 1/n^{t+1}$. }
{$Z$ --- a string on $m$ bits that is $n^{-\Omega(1)}$ close to uniform, with $m=t^{\Omega(1)}$.} {
Let $\alpha=\delta/4$. Let $\bitext$ be the one-bit extractor for non-oblivious bit-fixing source in Theorem~\ref{thm:bitext}. Let $G$ be the generating matrix of an asymptotically good $[r, m, d]_2$ code with $r=O(m) \leq n^{\alpha}$ and $d=\Omega(m)$. Thus $G$ is an $m \times r$ binary matrix.
} {alg:bfext}

\begin{enumerate}
\item Divide $X$ into $\ell=n^{\alpha}$ disjoint blocks, each with length $n^{1-\alpha}$.  
\item For each block $X_i, i \in [\ell]$, compute $Y_i=\bitext(X_i)$.
\item Let $Y=(Y_1, \cdots, Y_r)$ be the binary vector in $\F^r_2$. Compute $Z=GY$ where all operations are in $\F_2$.
\end{enumerate}
\end{algorithmsub}

We have the following theorem.

\BT \label{thm:bfext}
There exists a constant $c$ such that for any constant $\delta>0$ and all $n \in \N$, there exists an explicit extractor $\bfext: \bits^n \to \bits^m$ such that for any $(q, t, \gamma)$ non-oblivious bit-fixing source $X$ on $n$ bits with $q \leq n^{1-\delta}$, $t \geq c \log^{21} n$ and $\gamma \leq 1/n^{t+1}$, we have that
\[|\bfext(X)-U_m| \leq \e,\]

where $m=t^{\Omega(1)}$ and $\e=n^{-\Omega(1)}$.
\ET

\begin{thmproof}
Let the set of ``bad" bits in $X$ be $Q$, and the rest of the ``good" bits be $\overline{Q}$. Thus $|Q|=q \leq n^{1-\delta}$. Therefore, any block $X_i$ forms a $(q, t, \gamma)$ non-oblivious bit-fixing source on $n'=n^{1-\alpha}=n^{1-\frac{\delta}{4}}$ bits. 

Note that $q \leq n^{1-\delta}< n^{(1-\frac{\delta}{4})(1-\frac{3\delta}{4})} =n'^{1-\frac{3\delta}{4}}$. Thus by Theorem~\ref{thm:bitext} we have that each $Y_i$ is $n^{\Omega(1)}$-close to uniform. Moreover since $q \leq n^{1-\delta}$ we have that $\mathbf{I}_{q, t, \gamma}(\bitext) \leq q/n'^{1-\frac{3\delta}{8}}< n^{-\frac{3\delta}{8}}$. 

For any $i \in [\ell]$, the above means that with probability at least $1-n^{-\frac{3\delta}{8}}$ over the fixing of $X_i \cap \overline{Q}$, we have that $Y_i$ is fixed regardless of what $X_i \cap Q$ is. Thus, it is also true that with probability at least $1-n^{-\frac{3\delta}{8}}$ over the fixing of $\overline{Q}$, we have that $Y_i$ is fixed regardless of what $Q$ is. By a union bound, with probability at least $1-n^{-\frac{3\delta}{8}}n^{\frac{\delta}{4}}=1-n^{-\frac{\delta}{8}}$ over the fixing of $\overline{Q}$, we have that $(Y_1, \cdots, Y_{\ell})$ is fixed and thus $Y=(Y_1, \cdots, Y_r)$ is also fixed. 

Now consider a different distribution $X'$ where the bits in $\overline{Q}$ have the same distribution as $X$, while the bits in $Q$ are fixed to $0$ independent of $\overline{Q}$. Let $Y'=(Y'_1, \cdots, Y'_r)$ and $Z'$ be computed from $X'$ using the same algorithm. Then, by the above argument, we have that 

\[|Y - Y'| \leq n^{-\frac{\delta}{8}} \text{ and thus also } |Z - Z'| \leq n^{-\frac{\delta}{8}}.\]

Now consider $X', Y', Z'$. Note that by Theorem~\ref{thm:twise} $X'$ is $n^t \gamma \leq 1/n$-close to a distribution where the bits in $\overline{Q}$ are truly $t$-wise independent, and the bits in $Q$ are fixed to $0$. Thus from now on we will think of $X'$ as this distribution, since this only adds at most $1/n$ to the error.

Let $X''$ be another distribution where the bits in $\overline{Q}$ are completely uniform and independent, and the bits in $Q$ are fixed to $0$. Let $Y'', Z''$ be the corresponding random variables obtained from $X''$ instead of $X'$.

Take any non-empty subset $S \subseteq [m]$, and consider the random variable $V'_S=\bigoplus_{i \in S} Z'_i$.

Note that

\[V'_S=\bigoplus_{i \in S} Z'_i=\bigoplus_{i \in S} G_i Y'=(\sum_{i \in S}G_i) Y', \]

where $G_i$ stands for the $i$'th row of the matrix $G$. Since $G$ is the generating matrix of a $[r,m, d]_2$ code, for any non-empty subset $S \subseteq [m]$, we have that $\sum_{i \in S}G_i$ is a codeword. Thus it has at least $d$ $1$'s. 

Now let $V''_S$ be the corresponding random variable obtained from $X''$. Note that by Theorem~\ref{thm:bitext} each $Y''_i$ is $n^{-\Omega(1)}$-close to uniform, and now the $\{Y''_i\}$'s are independent of each other (since they are functions applied to independent blocks of $X''$). Therefore by Lemma~\ref{lem:derror} we have that 

\[|E[V''_S]-1/2| \leq (n^{-\Omega(1)})^{d}=2^{-\Omega(m \log n)}.\] 

Moreover, observe that $V'_S$ is the parity of at most $r=O(m)$ $Y'_i$'s. Since parity on $r$ bits can be computed by a depth-2 $\mathsf{AC}^0$ circuit (i.e., a DNF or CNF) of size $2^{O(r)}=2^{O(m)}$, and every $Y'_i$ is computed by a depth-4 $\mathsf{AC}^0$ circuit with size $n^{O(1)}$, we have that $V'_S$ can be computed by a depth-6 $\mathsf{AC}^0$ circuit with size at most $2^{O(m)} \poly(n)$.

We choose $m=min\{n^{0.9\alpha}, \beta t^{\frac{1}{21}}\}$ for some small constant $0<\beta<1$, so that $m=t^{\Omega(1)}$ (since $t \leq n$) and $r=O(m) \leq n^{\alpha}$. Note that now the $\mathsf{AC}^0$ circuit size is at most $2^{O(m)} \poly(n)  =2^{O(m)}$ since $t^{\frac{1}{21}} =\Omega(\log n)$. Note that the bits in $Q$ are fixed to $0$, thus $V'_S$ is computed by a depth-6 $\mathsf{AC}^0$ circuit with inputs from $\overline{Q}$. 

Setting $\e=2^{-m}$ in Theorem~\ref{thm:ac0}, we see that to $\e$-fool a depth-6 $\mathsf{AC}^0$ circuit with size at most $2^{O(m)}$, it suffices to take $O(\log(2^{O(m)}))^{21}=O(m^{21})$-wise independent distributions. By setting $\beta$ to be small enough, we can make this number less than $t$. Since in $X'$, the bits in $\overline{Q}$ are $t$-wise independent, we have that

\[|E[V'_S]-E[V''_S]| \leq 2^{-m}.\]

Thus

\[|E[V'_S]-1/2| \leq 2^{-\Omega(m \log n)}+2^{-m} < 2^{1-m}.\]

Note that this holds for every non-empty subset $S \subseteq [m]$. Thus $Z$ is $\e'$-biased for linear tests with $\e' <2 \cdot 2^{1-m}=2^{2-m}$. By the Lemma~\ref{lem:vxor} we have that 

\[|Z' -U_m| \leq 2^{m/2}2^{2-m}=2^{-\Omega(m)}.\]

Adding back the errors, we have

\[|Z-U_m| \leq 2^{-\Omega(m)}+1/n+n^{-\frac{\delta}{8}}=n^{-\Omega(1)}.\]
\end{thmproof}

The following theorem is implicit in \cite{Li15} and explicit in \cite{CZ15}

\BT [\cite{Li15, CZ15}] \label{thm:reduce} There exist constants $\delta, c' >0$ such that for every $n, t \in \N$ there exists a polynomial time computable function $\mathsf{reduce}: \bits^n \times \bits^n \to \bits^N$ with $N=n^{O(1)}$ satisfying the following property: if $X, Y$ are two independent $(n, k)$ sources with $k \geq c' t^4 \log^2 n$, then 

\[\Pr_{y \sim Y} [\mathsf{reduce}(X, y) \text{ is a } (q, t, \gamma) \text{ non-oblivious bit-fixing source }] \geq 1-n^{-\omega(1)},\]
where $q=N^{1-\delta}$ and $\gamma=1/N^{t+1}$.
\ET

Together with Theorem~\ref{thm:bfext} this immediately implies the following theorem.

\BT \label{thm:twosource}
There exists a constant $C>0$ such that for all $n \in \N$, there exists a polynomial time computable function $\twext: \bits^n \times \bits^n \to \bits^m$ satisfying the following: if $X, Y$ are two independent $(n, k)$ sources with $k \geq \log^C n$, then 

\[|(\twext(X, Y), Y)-(U_m, Y)| \leq \e,\]
where $m=k^{\Omega(1)}$ and $\e=n^{-\Omega(1)}$.
\ET

\begin{thmproof}
We first use Theorem~\ref{thm:reduce} to obtain a $(q, t, \gamma)$ non-oblivious bit-fixing source $Z$ on $N=n^{O(1)}$ bits, with $q=N^{1-\delta}$ and $\gamma=1/N^{t+1}$. We then apply the extractor for such sources in Theorem~\ref{thm:bfext}. By choosing $C$ large enough we can ensure that $k \geq c' t^4 \log^2 n$ and $t \geq c \log^{21} n$ (e.g., take $C=87$). Thus we see that $t=k^{\Omega(1)}$.

Therefore, by Theorem~\ref{thm:bfext} the extractor can output $t^{\Omega(1)}=k^{\Omega(1)}$ bits with error $n^{-\Omega(1)}$. Since the reduction succeeds with probability $1-n^{-\omega(1)}$ over the fixing of $Y$, the extractor is also strong in $Y$ and the final error is $\e=n^{-\Omega(1)}+n^{-\omega(1)}=n^{-\Omega(1)}$.
\end{thmproof}

\section{Conclusions and Open Problems}\label{sec:conc}
Constructing explicit two-source extractors is a challenging problem. Through a long line of research, the recent breakthrough result of Chattopadhyay and Zuckerman \cite{CZ15} has finally brought us close to the optimal. In this paper we managed to improve the output length of the (strong) two-source extractors in \cite{CZ15} from $1$ to $k^{\Omega(1)}$, but the error remains $n^{-\Omega(1)}$. The most obvious open problems are to improve the output length (say to $\Omega(k)$) and the error (say to exponentially small).

Both of these two problems seem challenging and requiring new ideas. Specifically, the current approach is to first reduce two independent sources to a $(q, t, \gamma)$ non-oblivious bit-fixing source, and then apply a deterministic extractor to such sources. This reduction step (implicit in \cite{Li15} and explicit in \cite{CZ15}) inherently depend on the alternating extraction method developed in previous work on independent source extractors). As a result, this technique, together with the requirement that $\gamma \leq 1/n^{t+1}$, seems to imply that $t$ can be at most $k^{\alpha}$ for some constant $\alpha<1$. As $t$ can be viewed as the entropy of the $(q, t, \gamma)$ non-oblivious bit-fixing source, it seems that we can extract at most $t=k^{Omega(1)}$ bits.

Second, the extractor for non-oblivious bit-fixing source crucially depends on resilient functions, where the analysis is done by bounding the influence of a coalition of variables. If the non-oblivious bit-fixing source has length $n^{O(1)}$ (to ensure polynomial time computability), then even one bit can have influence $\Omega(\log n/n^{O(1)})$ by the result of Kahn, Kalai and Linial \cite{KKL88}. Therefore we cannot hope to get error $n^{-\omega(1)}$ through this approach. However, there is indeed one way to get smaller error. That is to increase the length of the non-oblivious bit-fixing source. Indeed, by increasing the length to $n^{\omega(1)}$ we can get error $n^{-\omega(1)}$, but then the time for computing the extractor will be $n^{\omega(1)}$ as well.

Finally, an interesting observation of our work is that actually the bias of the one bit extractor in \cite{CZ15} is not very important (in \cite{CZ15} it has bias $n^{-\Omega(1)}$). Indeed, even if it only has constant bias, after the step of using the generating matrix $G$, we can see that the XOR of $\Omega(m)$ copies will have bias $2^{-\Omega(m)}$. However, at this moment this observation doesn't seem to help improve the parameters.
  
\bibliographystyle{alpha}

\bibliography{refs}

\end{document}